\documentclass{IEEEtran}
\usepackage[dvipsnames]{xcolor}
\usepackage{cite}
\usepackage{amsmath,amssymb,amsfonts}
\usepackage{mathtools}
\usepackage{lipsum}
\usepackage{algorithmic}
\usepackage{graphicx}
\usepackage{textcomp}
\usepackage{booktabs}
\usepackage{esint}
\def\BibTeX{{\rm B\kern-.05em{\sc i\kern-.025em b}\kern-.08em
    T\kern-.1667em\lower.7ex\hbox{E}\kern-.125emX}}
\begin{document}
\title{Design of Rectangular Waveguide-fed Metasurfaces for Near-Field Shaping using a Coupled Dipole Model}
\author{Insang Yoo, \IEEEmembership{Member, IEEE}, Dong Hwan Min \IEEEmembership{Student Member, IEEE},\\ Thomas Fromenteze, and Okan Yurduseven \IEEEmembership{Senior Member, IEEE}
\thanks{I. Yoo is with the School of Electrical and Electronic Engineering, Yonsei University, Seoul, Korea (\textit{Corresponding Author: Insang Yoo}, e-mail: insang.yoo@yonsei.ac.kr). D. H. Min is with the School of Electrical and Electronic Engineering, Yonsei University, Seoul, Korea. T. Fromenteze is with the University of Limoges, CNRS, XLIM, UMR 7252, France and with the Institut Universitaire de France. O. Yurduseven is with the Center for Wireless Innovation, Institute of Electronics, Communication and Information Technology, Queen’s University Belfast,
BT3 9DT Belfast, U.K.}}


\maketitle

\begin{abstract}
We present the design of rectangular waveguide-excited metasurfaces for near-field shaping using a coupled dipole framework. Waveguide-fed metasurfaces are array-like radiating systems typically constructed from one or more waveguides loaded with a series of subwavelength metamaterial apertures that function as radiators. The use of subwavelength radiating elements distributed across the aperture enables electromagnetic field control with subwavelength precision, offering significant potential for near-field shaping. Leveraging these capabilities, we demonstrate that the near-field patterns of rectangular waveguide-fed metasurfaces can be tailored using the coupled dipole model, which accounts for mutual interactions between metamaterial radiating elements. The validity and effectiveness of the proposed approach are verified through full-wave simulations and experiments in the X-band.
\end{abstract}


\begin{IEEEkeywords}
Aperture antenna, Leaky-wave antenna, Metasurfaces, Metamaterials.
\end{IEEEkeywords}

\section{Introduction} \label{sec:introduction}

\IEEEPARstart{S}{}haped near-fields play a critical role in various wireless systems, including radio-frequency identification (RFID) \cite{buffi2010focused,chou2010design,chou2015subsystem}, wireless power transfer \cite{borgiotti1966maximum,smith2017wpt}, and imaging systems \cite{hunt2013metamaterial,fromenteze2017single,molaei2022development}. Recently, controlled near-fields have attracted increasing attention for their potential applications in next-generation wireless networks \cite{zhang2022beam,cui2022channel,zhang20236g} and simultaneous wireless information and power transfer (SWIPT) systems \cite{zhang2013mimo}. These developments highlight the growing need for robust design methods to realize prescribed near-field patterns for antenna platforms at hand.

Among various radiative systems, we consider waveguide-fed metasurfaces, which have emerged as promising physical platforms for controlling electromagnetic radiation \cite{smith2017analysis}. The metasurfaces consist of a waveguide loaded with subwavelength metamaterial apertures, allowing for the manipulation of electromagnetic fields at subwavelength scales. Such architecture provides design freedom for tailoring near- and far-field distributions \cite{hunt2013metamaterial,boyarsky2021electronically}, while simplifying hardware implementation by eliminating the need for complex feed networks. Moreover, dynamic reconfiguration can be achieved using low-power tunable components such as diodes or liquid crystals \cite{sleasman2016waveguide,boyarsky2021electronically}, making the metasurfaces highly attractive for applications in wireless power transfer \cite{smith2017wpt,gowda2018focusing}, computational imaging \cite{molaei2022fourier}, and 6G communication systems \cite{shlezinger2021dynamic}.

\begin{figure}[!b]
\centering
\includegraphics[width=3.5in]{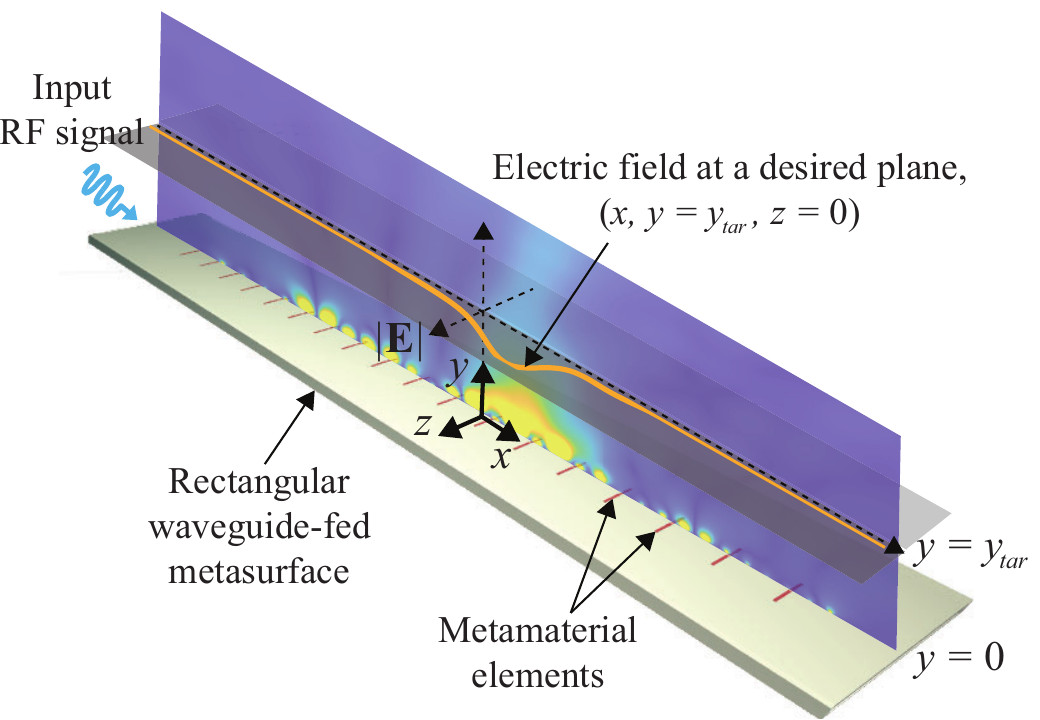}
\caption{Schematic of a rectangular waveguide-fed metasurface for near-field shaping. The geometry of individual metamaterial elements (e.g., rectilinear slots) is tuned to generate a desired field profile along the line defined by $\left(x,y=y_{tar},z=0\right)$.}
\label{Fig1_Schematic}
\end{figure}

Conventional approaches for designing waveguide-fed metasurfaces for near-field control often rely on holographic pattern synthesis methods \cite{smith2017wpt,yurduseven2017design}. While providing a systematic design framework, the holographic design method assumes weak coupling between the metamaterial elements and the guided modes, and neglects mutual interactions among neighboring radiators. As a result, the validity of the holographic approach may be reduced, and performance discrepancies can arise in metasurfaces whose elements are strongly coupled to the guided modes. Moreover, the inclusion of practical antenna parameters within holographic design frameworks necessitates extensive full-wave simulations and iterative optimization, thus reducing their efficacy and efficiency, particularly for the design of electrically large metasurface implementations.

To address these challenges, we present a design approach for waveguide-fed metasurfaces capable of generating sculpted near-field patterns using a coupled dipole model (CDM). The proposed method builds upon the dipole modeling framework \cite{pulido2017polarizability,yoo2022conformal}, which accounts for mutual interactions among metamaterial radiators. Focusing on rectangular waveguide-fed metasurfaces illustrated in Fig. \ref{Fig1_Schematic}, we integrate the CDM into an optimization framework and develop a systematic design method that incorporates practical antenna properties--such as S-parameters--into the metasurface design process. The validity and effectiveness of the proposed approach are verified through full-wave simulations and experiments, demonstrating that the proposed design method provides an efficient and accurate means for realizing waveguide-fed metasurfaces with tailored near-field patterns for various wireless applications.

\section{Design of Rectangular Waveguide-fed Metasurfaces for Near-field Shaping using Coupled Dipole Model} \label{theory}

We begin with a brief review of a CDM for rectangular waveguide-fed metasurfaces reported in \cite{pulido2016dda,yoo2022conformal}. Figure \ref{Fig1_Schematic} illustrates the schematic of a rectangular waveguide-fed metasurface, which consists of an array of subwavelength waveguide apertures--referred to as metamaterial radiators--embedded in the top wall of the waveguide. When the dimensions of the metamaterial radiator are much smaller than the guided wavelength, its scattered fields can be modeled as those by a set of polarizable dipoles, consisting of in-plane magnetic dipoles and an out-of-plane electric dipole \cite{bethe1944theory}. Such dipolar modeling of the radiators is valid for electrically small apertures of arbitrary geometry, including complementary metamaterial elements \cite{pulido2016dda,yoo2016efficient,yoo2022experimental}. The dipoles representing the radiator can be characterized with the effective polarizabilities, which can be determined either numerically \cite{pulido2017polarizability} or experimentally \cite{yoo2022experimental}.




Assuming $N_m$ metamaterial radiators are arranged along the centerline of the waveguide, the entire metasurface aperture can be effectively modeled as an array of dipoles excited by the guided modes, each consisting of a magnetic dipole oriented along $\hat{z}$ and an electric dipole oriented along $\hat{y}$. Under the coupled dipole framework, the analysis of the metasurface can be cast into a coupled matrix equation, given in \cite{pulido2016dda,yoo2022conformal} as
\begin{equation} \label{DDA_eq}
\begin{aligned}
\begin{bmatrix}
\mathbf{G}^{mm}_{zz} & \mathbf{G}^{me}_{zy} \\
\mathbf{G}^{em}_{yz} & \mathbf{G}^{ee}_{yy}
\end{bmatrix}
\begin{bmatrix}
\mathbf{m}_{z} \\
\mathbf{p}_{y}
\end{bmatrix}
=\begin{bmatrix}
\mathbf{H}^{i}_{z} \\
\mathbf{E}^{i}_{y}
\end{bmatrix},
\end{aligned}
\end{equation}
with $\mathbf{m}_{z},\mathbf{p}_{y}\in\mathbb{C}^{N_{m} \times 1}$ being the magnetic and electric dipole moments representing the metamaterial elements, respectively. $\mathbf{H}^{i}_{z}$ and $\mathbf{E}^{i}_{y}$ represent the magnetic and electric fields incident on the element, respectively. The off-diagonal components of $\mathbf{G}^{}_{}\in\mathbb{C}^{N_{m} \times N_{m}}$ matrices are the sum of the Green's functions in the rectangular waveguide and the free-space Green's functions. The Green's functions are used to capture the mutual interaction between the metamaterial radiators. The diagonal entries of $\mathbf{G}^{mm}_{zz}$ and $\mathbf{G}^{ee}_{yy}$ are the inverse of the polarizabilities, while those of $\mathbf{G}^{me}_{zy}$ and $\mathbf{G}^{em}_{yz}$ are zeros. By solving the matrix equation in (\ref{DDA_eq}) for the dipole moments, the near-field pattern can be computed as the superposition of the dipole contributions obtained using the free-space Green’s functions. It is worth noting that the model in (\ref{DDA_eq}) accurately captures the dipolar coupling between metamaterial radiators through their scattered fields. Accordingly, the design approach presented in this work is distinct from conventional holographic pattern synthesis methods, whose validity is limited only to metasurfaces with radiators that are weakly coupled to the guided modes \cite{smith2017wpt, yurduseven2017design}.

\begin{figure}[!t]
\centering
\includegraphics[width=3.5in]{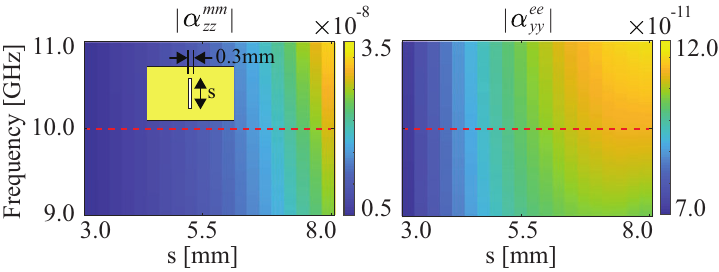}
\caption{The extracted effective magnetic polarizability (left) and electric (right) polarizability as functions of frequency and slot length $s$ (inset).}
\label{Fig2_Polarizability}
\end{figure}


Given the dipole model in (\ref{DDA_eq}), we design metamaterial radiators for rectangular waveguide-fed metasurfaces aimed at shaping near-field patterns. For the design, we consider a rectilinear transverse slot inserted into the top wall of a rectangular waveguide for structural simplicity. Although such slots are non-resonant and provide limited phase weights, we find that, when placed with subwavelength spacing, the slots are sufficient to realize the prescribed near-field patterns considered in this work. The waveguide is filled with Rogers 4003C substrate ($\epsilon_r = 3.55$, $\tan\delta = 0.0027$) and has a width of $a = 11.07$ mm and a height of $b = 1.52$ mm, selected to support propagation of the dominant TE${_{10}}$ mode within X-band. The slot width is fixed at $0.3$ mm ($\sim 0.01\lambda_{0}$ at $10.0$ GHz), and the polarizability extraction method is applied over a swept slot length ranging from $3.0$ mm to $8.0$ mm ($\sim 0.1\lambda_{0}$ and 0.27$\lambda_{0}$ at $10.0$ GHz) to obtain the effective polarizabilities as functions of frequency and slot length. The polarizabilities are computed using the scattering parameters as $\alpha_{yy}^{ee} = \frac{jab\beta_{}}{2k^{2}}\left(S_{21}+S_{11}-1\right)$ and 
$\alpha_{zz}^{mm} = \frac{jab}{2\beta_{}}\left(S_{21}-S_{11}-1\right)$, where $\beta$ and $k$ represent the propagation constant of the TE$_{10}$ mode and wavenumber in the substrate, respectively \cite{pulido2017polarizability}. The computed polarizabilities are shown in Fig. \ref{Fig2_Polarizability}, illustrating that the amplitude of the magnetic polarizability increases with slot length and frequency.


Next, we use the dipole model in (\ref{DDA_eq}) as a forward model for the metasurface by incorporating the polarizabilities shown in Fig. \ref{Fig2_Polarizability} into (\ref{DDA_eq}). Assuming $N_m = 30$ slots spaced at $8.5$ mm ($\sim 0.28\lambda_{0}$ at $10.0$ GHz), we employ a surrogate optimization method \cite{queipo2005surrogate} to determine the set of slot lengths that minimize a defined cost function. Note that the spacing was chosen to be subwavelength to utilize the phase advance of the guided modes and to ensure the validity of the dipole model. A closer spacing may lead to errors in predicting the near-field patterns of the metasurface due to coupling between the slots through higher-order modes \cite{pulido2017polarizability}, which is not considered in (\ref{DDA_eq}).

We design three metasurface examples by considering the cost functions and parameters listed in Table \ref{table_cost_function}. The functions $c_{nf,k}$ and $c_{sp,k}$ in Table \ref{table_cost_function} measure, respectively, the difference between the target and generated near-field patterns at the reference plane $y = y_{ref}$, and the sum of the scattering parameters, given by
\begin{equation} \label{cost_fun1}
\begin{aligned}
c_{nf,k} &= -\frac{1}{N_{x}}\sum_{i=1}^{N_{x}}\big\{E_{x}^{des} - E_{x}(f_{op,k})\big\}, \\
c_{sp,k} &= -\Big(|S_{11}(f_{op,k})| + |S_{21}(f_{op,k})|\Big),
\end{aligned}
\end{equation}
where $N_x$ denotes the number of sampling points along the line on the target plane, i.e., $(x, y = y_{ref}, z = 0)$. Additionally, the functions used to measure the ratio of peak amplitudes are defined as
\begin{equation} \label{cost_fun2}
\begin{aligned}
c_{peak} &= \frac{\text{max}\{E_{x,peak1}^{}\left(f_{op}\right),E_{x,peak2}^{}\left(f_{op}\right)\}}{\text{min}\{E_{x,peak1}^{}\left(f_{op}\right),E_{x,peak2}^{}\left(f_{op}\right)\}}, \\
c_{peak,k} &= \frac{\underset{k}{\text{max}}\{E_{x,peak}^{}\left(f_{op,k}\right)\}}{\underset{k}{\text{min}}\{E_{x,peak}^{}\left(f_{op,k}\right)\}},
\end{aligned}
\end{equation}
with $E_{x,peak}$ is the peak value of the electric field along the line on the target plane. The weights in the cost functions in Table \ref{table_cost_function} are chosen empirically.

\begin{table*}[t] 
    \centering 
    \caption{Cost Functions and Parameters used for Metasurface Design. $f$ is Gaussian function.}
    \label{tab:wide_table}
    \begin{tabular}{|c|c|c|c|}
        \hline
        Type & \vtop{\hbox{\strut Equal-amplitude}\hbox{\strut Two Focused Beams $(k=1)$}} & \vtop{\hbox{\strut Different-amplitude}\hbox{\strut Two Focused Beams $(k=1)$}} & \vtop{\hbox{\strut Frequency-scanned Beams}\hbox{\strut $(k=1,2,3)$}} \\
        \hline
        Frequency (GHz) & $f_{op} =10.0$ & $f_{op} =10.0$ & $f_{op,1} =8.5$, $f_{op,2} =9.5$, $f_{op,3} =10.5$ \\
        \hline
         Cost Function & $c_{nf,k} + 0.05 c_{sp,k} + 0.10 c_{peak}$ & $c_{nf,k} + 0.05 c_{sp,k}$ & $c_{nf,k} + 0.10 c_{sp,k} +0.05 c_{peak,k}$ \\
        \hline
         Near-field Pattern ($E_{x}^{des}$) & $f\left(x,\mu_{1},\sigma_{1}\right)+f\left(x,\mu_{2},\sigma_{2}\right)$ & $\frac{1}{2}f\left(x,\mu_{1},\sigma_{1}\right)+f\left(x,\mu_{2},\sigma_{2}\right)$ & $f\left(x,\mu_{k},\sigma_{k}\right)$ \\
        \hline
         Parameters (mm) &  \vtop{\hbox{\strut $\mu_{1}=-40$, $\mu_{2}=30$,}\hbox{\strut $\sigma_{1}=\sigma_{2}=13$}} & \vtop{\hbox{\strut $\mu_{1}=-40$, $\mu_{2}=30$,}\hbox{\strut $\sigma_{1}=\sigma_{2}=13$}} & \vtop{\hbox{\strut $\mu_{1}=-\mu_{3}=-50, \mu_{2}=0$,}\hbox{\strut $\sigma_{1}=\sigma_{2}=\sigma_{3}=13$}} \\
        \hline
        Target Plane (mm) & $y_{tar}=90$ & $y_{tar}=75$ & $y_{tar}=85$ \\
        \hline
    \end{tabular}
    \label{table_cost_function}
\end{table*}

\section{Experimental Demonstration} \label{design}

For experimental verification of the designed metasurfaces, the rectangular waveguide described in Section \ref{theory} was implemented with an equivalent substrate-integrated waveguide (SIW) due to its compact form factor and low leakage. The equivalent SIW was designed through full-wave simulations of both the rectangular waveguide and the SIW with identical lengths. The width of the SIW was adjusted so that the phase of $S_{21}$ matched that of the rectangular waveguide. The via diameter ($D_v = 1.0$ mm) and spacing ($s_v = 2.0$ mm) were chosen to minimize losses in the designed SIW \cite{bozzi2011review}.

\begin{figure}[!b]
\centering
\includegraphics[width=3.5in]{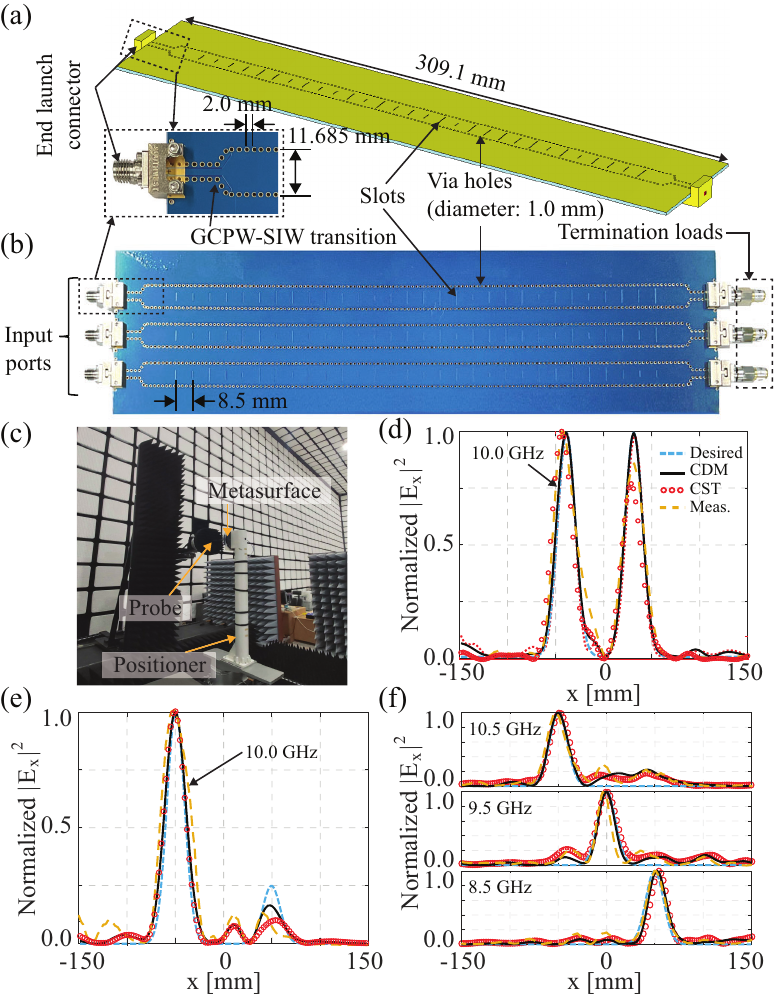}
\caption{(a) Schematic of the designed metasurface example generating two beams at the target plane, realized with the substrate-integrated waveguide and grounded coplanar waveguide-to-SIW transition feeds. (b) Photo of the fabricated metasurface examples, each with $N_{m}=30$ rectangular slot radiators (photo edited for compact illustration). The metasurfaces are designed to generate two near-field focused beams with the same and different amplitudes (top, middle rows), and frequency-scanned beams (bottom row). (c) Measurement setup using a planar near-field scan system. The cross-sectional plots of the normalized squared electric field along the line $y=y_{ref}$ for (d) the equal-amplitude design, (e) the different peak-amplitude design, and (f) the frequency-scanned beam design.}
\label{Fig3_photo}
\end{figure}

A grounded coplanar waveguide (GCPW)-to-SIW transition structure was then designed and connected to an end-launch connector to feed the SIW. An identical transition structure, terminated with a $50 \Omega$ load, was attached to the opposite end of the SIW for matched termination. The slots, with lengths determined in Section \ref{theory}, were subsequently etched into the top wall of the SIW to form the SIW-fed metasurface, as illustrated in Fig.~\ref{Fig3_photo}(a). It should be noted that the GCPW-to-SIW transitions include apertures, which were etched into the bottom wall of the SIW to prevent fields leaking through them from contributing to the overall near-field pattern. We then analyzed the designed metasurfaces using a full-wave electromagnetic solver. 


Next, we fabricated the metasurfaces designed to generate two focused beams with equal amplitudes, two focused beams with different amplitudes, and frequency-scanned beams, as shown in Fig. \ref{Fig3_photo}(b). The fabricated metasurfaces were characterized using a planar near-field scan system and a probe antenna. During the measurements, the near-field data were sampled on a regular grid over a plane parallel to the metasurface aperture (i.e., the $xz$-plane shown in Fig. \ref{Fig1_Schematic}), located $150$ mm away from the aperture, as illustrated in Fig. \ref{Fig3_photo}(c). The measured data were then processed to reconstruct the near-field distribution on the target plane ($y = y_{ref}$) using the holographic back-projection technique \cite{balanis2005antenna}.

Figure \ref{Fig3_photo}(d)-(f) shows the cross-sectional plots of the normalized, squared electric field of the fabricated metasurface samples, shown in Fig. \ref{Fig3_photo}(b). Overall, great agreement between the simulated, measured near-field patterns and those predicted by the dipole model. Specifically, for the metasurface designed to generate two beams with equal peak amplitudes, two beams were formed near the target positions (i.e., $x=-40$ mm and $30$ mm) at the operating frequency of $10.0$ GHz, demonstrating very close agreement between the simulated and measured patterns and those predicted by the dipole model--although the effects of the GCPW transitions were not considered in the dipole model. The reduced amplitude of the peak at $x=30$ mm observed in both the simulated and measured patterns in Fig. \ref{Fig3_photo}(d) is attributed to increased losses in the fabricated SIW structure. As shown in Fig. \ref{Fig3_photo}(e), for the metasurface designed to generate two beams with different peak amplitudes, two beams were formed near the target positions (i.e., $x=-40$ mm and $30$ mm) at the operating frequency of $10.0$ GHz in both the simulated and predicted near-field patterns. The intended peak ratio was $E_{x,peak1} / E_{x,peak2} = 0.5$. In the measured pattern shown in Fig. \ref{Fig3_photo}(e), two peaks were also observed, although one peak was slightly shifted from its target position at $x=30$ mm. For the metasurface designed to exhibit frequency-scanned beams, the electric field peaks appear at the target positions (i.e., $x = 50$, $0$, and $-50$ mm) at $8.5$, $9.5$, and $10.5$ GHz, respectively, as shown in Fig. \ref{Fig3_photo}(f).




We also measured the S-parameters of the fabricated metasurface examples shown in Fig. \ref{Fig3_photo}(b). For the two equal-amplitude beam design, $|S_{11}|$ exhibits a sharp dip near the operating frequency of $10.0$ GHz, with measured and simulated values of $–8.2$ dB and $–10.7$ dB, respectively. For the two different-amplitude beam design, a $|S_{11}|$ dip is also observed at $10.0$ GHz, with measured and simulated values of $–8.7$ dB and $–16.0$ dB, respectively. For the frequency-swept beam design, $|S_{11}|$ dips occur at the operating frequencies of $8.5$, $9.5$, and $10.5$ GHz. The measured $|S_{11}|$ values are $–12.9$, $–6.8$, and $–10.1$ dB, while the simulated values are $–14.2$, $–8.6$, and $–8.2$ dB, respectively. Note that for the frequency-swept beam design, the ratio of the peak values was considered, as in (\ref{cost_fun2}), leading to different levels of $|S_{11}|$. Slight frequency shifts in the measured $|S_{11}|$ dips were also observed, which could be mitigated by accurate characterization of the substrate materials used in the design. For all metasurface examples, $|S_{21}|$ at the operating frequencies was measured and kept below $-15.0$ dB.

Note that the dipole model employed for the metasurface design assumes a rectangular waveguide—-rather than a SIW with vias—-and does not account for the GCPW-to-SIW transition structure. Incorporating the effects of the transition into the dipole model remains an important step for future work; therefore, S-parameters predicted by the dipole model are not included. Also, note that metamaterial radiators strongly coupled to the guided modes--such as the resonant elements reported in \cite{pulido2017polarizability,yoo2016efficient}--can be employed to achieve a broader phase response and greater degrees of freedom for shaping complex near-field distributions. The primary goal of this work, however, is to demonstrate the validity of the proposed design approach for waveguide-fed metasurfaces that generate prescribed near-field patterns using the CDM. However, the use of resonant elements will be essential for developing high-performance metasurfaces capable of producing more complex near-field patterns and thus remains an important direction for future research.

\begin{figure}[!t]
\centering
\includegraphics[width=2.0in]{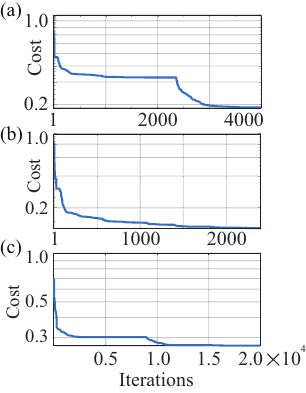}
\caption{The evolution of the cost function for (a) the equal-amplitude design, (b) the different peak-amplitude design, and (c) the frequency-scanned beam design.}
\label{Fig4_sparameters}
\end{figure}


To demonstrate the effectiveness of the proposed design method, the evolution of the cost function as a function of iteration number for the designed metasurface examples is shown in Figs. \ref{Fig4_sparameters}(a)–(c). The stopping criterion for the surrogate optimization was set to the maximum number of iterations--$4000$, $2400$, and $20000$, respectively--to ensure convergence. As shown in Figs. \ref{Fig4_sparameters}(a)-(c), the cost functions exhibit several abrupt decreases during the early stages and subsequently converge rapidly, confirming the efficiency of the design approach. It is also noted that the evolution and final value of the cost function may vary depending on the desired near-field distribution and the definition of the cost function, thus indicating that the appropriate selection of the cost function and stopping criteria is essential for the design.

For better illustration, it should be emphasized that the design method using the CDM offers a significant computational advantage over the full-wave simulations. For the equal-amplitude beam design, the CDM implemented in MATLAB on a workstation (64-bit, 3.19 GHz CPU) requires only $0.02$ seconds per frequency, whereas a full-wave simulation of the metasurface--with an array of $30$ slot elements embedded in the rectangular waveguide wall and discretized into $1.9$ million cells--takes $892$ seconds per frequency.

\section{Conclusion}

We have presented the design of rectangular waveguide–fed metasurfaces with tailored near-field patterns using the CDM and numerical optimization method. The validity of the proposed design approach has been verified through full-wave numerical simulations and experimental demonstrations using fabricated SIW-fed metasurfaces, which employ equivalent SIW structures and rectilinear slots. Given the results, the proposed method, together with the dipole model for metasurfaces, can be extended to the design and analysis of metasurfaces for various wireless systems, including computational imaging, wireless communication, and wireless power transfer systems.

\section*{Acknowledgment}
This work was partially supported by the Yonsei University Research Fund (Grant No.: 2024-22-0062). This work was also supported partially by the National Research Foundation of Korea (NRF) grant funded by the Ministry of Science and ICT (MSIT) (Grant No.: RS-2024-00512014 (Korea-EU Cooperation Program), RS-2025-00554900).

\bibliographystyle{IEEEtran}

\begin{thebibliography}{10}
\providecommand{\url}[1]{#1}
\csname url@samestyle\endcsname
\providecommand{\newblock}{\relax}
\providecommand{\bibinfo}[2]{#2}
\providecommand{\BIBentrySTDinterwordspacing}{\spaceskip=0pt\relax}
\providecommand{\BIBentryALTinterwordstretchfactor}{4}
\providecommand{\BIBentryALTinterwordspacing}{\spaceskip=\fontdimen2\font plus
\BIBentryALTinterwordstretchfactor\fontdimen3\font minus \fontdimen4\font\relax}
\providecommand{\BIBforeignlanguage}[2]{{%
\expandafter\ifx\csname l@#1\endcsname\relax
\typeout{** WARNING: IEEEtran.bst: No hyphenation pattern has been}%
\typeout{** loaded for the language `#1'. Using the pattern for}%
\typeout{** the default language instead.}%
\else
\language=\csname l@#1\endcsname
\fi
#2}}
\providecommand{\BIBdecl}{\relax}
\BIBdecl

\bibitem{buffi2010focused}
A.~Buffi, A.~Serra, P.~Nepa, G.~Manara \emph{et~al.}, ``A focused planar microstrip array for 2.4 ghz rfid readers,'' \emph{IEEE Trans. Ant. Propag.}, vol.~58, no.~5, pp. 1536--1544, 2010.

\bibitem{chou2010design}
H.-T. Chou, T.-M. Hung, N.-N. Wang, H.-H. Chou, C.~Tung, and P.~Nepa, ``Design of a near-field focused reflectarray antenna for 2.4 ghz rfid reader applications,'' \emph{IEEE Trans. Ant. Propag.}, vol.~59, no.~3, pp. 1013--1018, 2010.

\bibitem{chou2015subsystem}
H.-T. Chou, M.-Y. Lee, and C.-T. Yu, ``Subsystem of phased array antennas with adaptive beam steering in the near-field rfid applications,'' \emph{IEEE Ant. Wireless Propag. Lett.}, vol.~14, pp. 1746--1749, 2015.

\bibitem{borgiotti1966maximum}
G.~v. Borgiotti, ``Maximum power transfer between two planar apertures in the fresnel zone,'' \emph{IEEE Trans. Ant. Propag.}, vol.~14, no.~2, pp. 158--163, 1966.

\bibitem{smith2017wpt}
D.~R. Smith, V.~R. Gowda, O.~Yurduseven, S.~Larouche, G.~Lipworth, Y.~Urzhumov, and M.~S. Reynolds, ``An analysis of beamed wireless power transfer in the {F}resnel zone using a dynamic, metasurface aperture,'' \emph{J. Appl. Phys.}, vol. 121, no.~1, p. 014901, 2017.

\bibitem{hunt2013metamaterial}
J.~Hunt, T.~Driscoll, A.~Mrozack, G.~Lipworth, M.~Reynolds, D.~Brady, and D.~R. Smith, ``Metamaterial apertures for computational imaging,'' \emph{Science}, vol. 339, no. 6117, pp. 310--313, 2013.

\bibitem{fromenteze2017single}
T.~Fromenteze, M.~Boyarsky, J.~Gollub, T.~Sleasman, M.~Imani, and D.~R. Smith, ``Single-frequency near-field mimo imaging,'' in \emph{Proc. 11th Eur. Conf. Antennas Propag. (EUCAP)}.\hskip 1em plus 0.5em minus 0.4em\relax IEEE, 2017, pp. 1415--1418.

\bibitem{molaei2022development}
A.~M. Molaei, T.~Fromenteze, V.~Skouroliakou, T.~V. Hoang, R.~Kumar, V.~Fusco, and O.~Yurduseven, ``Development of fast fourier-compatible image reconstruction for 3d near-field bistatic microwave imaging with dynamic metasurface antennas,'' \emph{IEEE Trans. Veh. Technol.}, vol.~71, no.~12, pp. 13\,077--13\,090, 2022.

\bibitem{zhang2022beam}
H.~Zhang, N.~Shlezinger, F.~Guidi, D.~Dardari, M.~F. Imani, and Y.~C. Eldar, ``Beam focusing for near-field multiuser mimo communications,'' \emph{IEEE Trans. Wireless Commun.}, vol.~21, no.~9, pp. 7476--7490, 2022.

\bibitem{cui2022channel}
M.~Cui and L.~Dai, ``Channel estimation for extremely large-scale mimo: Far-field or near-field?'' \emph{IEEE Trans. Commun.}, vol.~70, no.~4, pp. 2663--2677, 2022.

\bibitem{zhang20236g}
H.~Zhang, N.~Shlezinger, F.~Guidi, D.~Dardari, and Y.~C. Eldar, ``6g wireless communications: From far-field beam steering to near-field beam focusing,'' \emph{IEEE Commun. Mag.}, vol.~61, no.~4, pp. 72--77, 2023.

\bibitem{zhang2013mimo}
R.~Zhang and C.~K. Ho, ``Mimo broadcasting for simultaneous wireless information and power transfer,'' \emph{IEEE Trans. Wireless Commun.}, vol.~12, no.~5, pp. 1989--2001, 2013.

\bibitem{smith2017analysis}
D.~R. Smith, O.~Yurduseven, L.~P. Mancera, P.~Bowen, and N.~B. Kundtz, ``Analysis of a waveguide-fed metasurface antenna,'' \emph{Phys. Rev. Appl.}, vol.~8, no.~5, p. 054048, 2017.

\bibitem{boyarsky2021electronically}
M.~Boyarsky, T.~Sleasman, M.~F. Imani, J.~N. Gollub, and D.~R. Smith, ``Electronically steered metasurface antenna,'' \emph{Sci. Rep.}, vol.~11, no.~1, pp. 1--10, 2021.

\bibitem{sleasman2016waveguide}
T.~Sleasman, M.~F. Imani, W.~Xu, J.~Hunt, T.~Driscoll, M.~S. Reynolds, and D.~R. Smith, ``Waveguide-fed tunable metamaterial element for dynamic apertures,'' \emph{IEEE Trans. Antennas Propag.}, vol.~15, pp. 606--609, 2016.

\bibitem{gowda2018focusing}
V.~R. Gowda, M.~F. Imani, T.~Sleasman, O.~Yurduseven, and D.~R. Smith, ``Focusing microwaves in the fresnel zone with a cavity-backed holographic metasurface,'' \emph{IEEE Access}, vol.~6, pp. 12\,815--12\,824, 2018.

\bibitem{molaei2022fourier}
A.~M. Molaei, T.~Fromenteze, S.~Hu, V.~Fusco, and O.~Yurduseven, ``Fourier-based near-field three-dimensional image reconstruction in a multistatic imaging structure using dynamic metasurface antennas,'' \emph{IEEE Trans. Comput. Imag.}, vol.~8, pp. 1089--1100, 2022.

\bibitem{shlezinger2021dynamic}
N.~Shlezinger, G.~C. Alexandropoulos, M.~F. Imani, Y.~C. Eldar, and D.~R. Smith, ``Dynamic metasurface antennas for 6g extreme massive mimo communications,'' \emph{IEEE Wireless Commun.}, vol.~28, no.~2, pp. 106--113, 2021.

\bibitem{yurduseven2017design}
O.~Yurduseven, D.~L. Marks, J.~N. Gollub, and D.~R. Smith, ``Design and analysis of a reconfigurable holographic metasurface aperture for dynamic focusing in the fresnel zone,'' \emph{IEEE Access}, vol.~5, pp. 15\,055--15\,065, 2017.

\bibitem{pulido2017polarizability}
L.~Pulido-Mancera, P.~T. Bowen, M.~F. Imani, N.~Kundtz, and D.~Smith, ``Polarizability extraction of complementary metamaterial elements in waveguides for aperture modeling,'' \emph{Phys. Rev. B}, vol.~96, no.~23, p. 235402, 2017.

\bibitem{yoo2022conformal}
I.~Yoo and D.~R. Smith, ``Design of conformal array of rectangular waveguide-fed metasurfaces,'' \emph{IEEE Trans. Antennas Propag.}, vol.~70, no.~7, pp. 6060--6065, 2022.

\bibitem{pulido2016dda}
L.~Pulido-Mancera, T.~Zvolensky, M.~F. Imani, P.~T. Bowen, M.~Valayil, and D.~R. Smith, ``Discrete dipole approximation applied to highly directive slotted waveguide antennas,'' \emph{IEEE Antennas Wireless Propag. Lett.}, vol.~15, pp. 1823--1826, 2016.

\bibitem{bethe1944theory}
H.~A. Bethe, ``Theory of diffraction by small holes,'' \emph{Phys. Rev.}, vol.~66, no. 7-8, p. 163, 1944.

\bibitem{yoo2016efficient}
I.~Yoo, M.~F. Imani, T.~Sleasman, and D.~R. Smith, ``Efficient complementary metamaterial element for waveguide-fed metasurface antennas,'' \emph{Opt. Exp.}, vol.~24, no.~25, pp. 28\,686--28\,692, 2016.

\bibitem{yoo2022experimental}
I.~Yoo, D.~R. Smith, and M.~Boyarsky, ``Experimental characterization of a waveguide-fed varactor-tuned metamaterial element using the coupled dipole framework,'' \emph{IEEE Antennas Wireless Propag. Lett.}, vol.~22, no.~2, pp. 387--391, 2022.

\bibitem{queipo2005surrogate}
N.~V. Queipo, R.~T. Haftka, W.~Shyy, T.~Goel, R.~Vaidyanathan, and P.~K. Tucker, ``Surrogate-based analysis and optimization,'' \emph{Prog. Aerosp. Sci.}, vol.~41, no.~1, pp. 1--28, 2005.

\bibitem{bozzi2011review}
M.~Bozzi, A.~Georgiadis, and K.~Wu, ``Review of substrate-integrated waveguide circuits and antennas,'' \emph{IET Microw., Antennas Propag.}, vol.~5, no.~8, pp. 909--920, 2011.

\bibitem{balanis2005antenna}
C.~A. Balanis, \emph{Antenna Theory: Analysis and Design}.\hskip 1em plus 0.5em minus 0.4em\relax Wiley, 2005.

\end{thebibliography}

\end{document}